\documentclass[aps,prl,twocolumn,groupedaddress,showpacs]{revtex4-1}
\usepackage{amsmath}
\usepackage{graphicx}
\usepackage{epstopdf}
\usepackage{epsfig}
\usepackage{amssymb}
\usepackage{bm}
\usepackage{color}

\renewcommand{\thefootnote}{\fnsymbol{footnote}}

\begin{document}

\title{Commuting Matrix Solutions of PQCD Evolution Equations}
\pacs{12.38.Bx}
\author{Mehrdad Goshtasbpour$^{1}$ and Seyed Ali Shafiei$^{2}$
\linebreak}
\affiliation{$^{1}$ Dept. of Physics, Shahid Beheshti University, G. C., Evin 19834, Tehran, Iran. \\ $^{2}$ Biophysics-Biochemistry Department, Rafsanjan University of Medical Sciences (RUMS), Rafsanjan, Iran.} 

\begin{scriptsize}
\end{scriptsize}

\begin{abstract}
 A method of obtaining parton distributions directly from data is revealed in this series. In the process, the first step would be developing appropriate matrix solutions of the evolution equations in \textit{x} space. A division into commuting and non-commuting matrix solutions has been made. Here, well-developed commuting matrix solutions are presented. Results for finite LO evolution match those of standard LO sets. There is a real potential of doing non-parametric data analysis. \\
\end{abstract}
 
\maketitle

\section{Introduction}

Given conditions on parton distribution functions (pdfs), coming from existing measurements in the accelerators, e.g. from a measurement of the structure function ‎$ {F_2} $‎ in DIS, through a fixed order (LO, NLO, ...) formula from PQCD for $ {F_2} $ , or a given cross section for pp scattering in a similar manner, can be considered the initial conditions for the pdfs in the intgro-differential DGLAP evolution equations of the same order in perturbation; based on which solving the DGLAP equations generally allows PQCD prediction for processes of interest in accelerators, and eventually in the colliders. Thus, the method of solution of DGLAP proposed in the year 2000 \cite{Spin2000}, and beginning to be presented in details here,\thefootnote{*} can be of serious interest. For simplicity the method is illustrated at LO PQCD.

DGLAP evolution equation is of the following simple 
form for the ‎\textit{non-singlet}‎ quark distribution ‎$ f_{NS} $:
\begin{equation}\label{1-1}
\frac{{\partial {f_{NS}}(x,t)}}{{\partial t}} = \int_x^1 {\frac{{dy}}{y}} {P_{qq}}(\frac{x}{y}){f_{NS}}(y,t).
\end{equation}

Usually, the derivative in (\ref{1-1}) is with respect to ‎$ Ln({Q^2})$, where
\begin{equation}\label{1-3}
 ‎\frac{dt}{d Ln(Q^2)}=‎\frac{\alpha‎_s‎(Q^2)}{2\pi}.‎
\end{equation}

In (\ref{1-3}), and throughout this work, the LO value of the strong coupling constant ${\alpha _s}({Q^2})$ is used, either in the ‘historical approach’ based on $\Lambda _{QCD}^2$, or
in the present day approach based on experimental value of ${\alpha _s}({M_z})$ directly, \cite{Stirling, MSTW}, equivalently (with no difference).

\section{$n$ dimensional space, pdf vectors, splitting functions matrices}
Define discrete Bjorken ‎$ x_i $,
\begin{eqnarray}\label{2-1}
 {x_i} = {({x_1})^i}, i = 1,...,n
 , 
 \end{eqnarray}
for large ‎$ x_1< 1$. A set of ‎\textit{n}‎ basis vectors for each pdf set can be defined on each $ x_i $. The splitting function, in terms of (\ref{2-1}), can be calculated as an ‎$ n \times n $‎ matrix. The choice of $x_1$ and $n$ in (\ref{2-1}) is eventually constrained  by a close fit to the ‎\textit{x}‎ points of the data, for better interpolation.\\

\subsection{Convolution integrals}
Convolution integrals are of the form\thefootnote{**}: 
\begin{equation}\label{2-2}
P \otimes f \equiv \int_x^1 {\frac{{dy}}{y}} P(\frac{x}{y})f(y,t) = \int_x^1 {\frac{{dy}}{y}} P(y)f(\frac{x}{y},t),
\end{equation}
in which the simplest parton distribution function ‎$f(x,t)={f_{NS}}(x,t) $‎ in the discrete basis is an $n$-tuple,  whose $i$th  element is ‎$ f({x_i},t) $.  As  the general convolution integral is made a discrete sum, ${P_{ij}},i,j = q,g$ within the kernel $P$, see (\ref{1-4}), become lower triangular (l.t.), due to the limits of integration of the resulting sum, and banded (bdd), due to divisions of ${x_i}$ in (‎\ref{2-1}), see (\ref{2-5}) to (\ref{2-8}). The resulting algebraic simplicity, i.e, commutativity, of l.t. bdd matrices is the essential element of the following analysis of the solutions of DGLAP equations.

\subsection{A hybrid finite difference computation of splitting functions, an example}  
To calculate the matrix form of a splitting function, e.g.‎ $ {P_{qq}} $‎  in ‎(\ref{1-1})‎, a combination of discrete finite difference approximation of the unknown pdfs and continuous integration of the known splitting functions is mixed in our method of evaluation of the convolution, as follows. Then, ${({P_{qq}})_{ik}}$   is extracted from the coefficients of ‎$ {f_k} = f({x_k},t) $‎. At LO, ‎$ {P_{qq}} $ is given in the following equivalent forms \cite{Altarelli}:\\ 
\begin{equation}\label{2-3}
{P_{qq}} = \frac{4}{3}{ \bigg [\frac{{1 + {y^2}}}{{1 - y}} \bigg ]_ + } = \frac{4}{3}\bigg[\frac{{1 + {y^2}}}{{{{(1 - y)}_ + }}} + \frac{3}{2}\delta (1 - y)\bigg].
\end{equation}
Given the definition of "+" regularization, aimed at removal of the infinity at $y = 1$ in the kernel ${P_{qq}}$ in the context of the convolution integral, 
\begin{eqnarray}
\begin{gathered}
\int_{z}^{1}dx f(x) \bigg [‎\frac{g(x)}{1-x}‎ \bigg ]_+=\hfill \\ ‎\hfill‎ \\
\int_{z}^{1}dx ‎\frac{f(x)-f(1)}{1-x}g(x)‎-f(1)\int_{0}^{z}dx ‎\frac{g(x)}{1-x},‎
\end{gathered}
\label{2-4}‎
\end{eqnarray}
convolution in (‎\ref{1-1})‎ becomes:
\begin{eqnarray}
\begin{gathered}
 I =\int_x^1 {\frac{{dy}}{y}y.}f(\frac{x}{y}). \frac{4}{3}{\bigg(\frac{{1 + {y^2}}}{{1 - y}}\bigg)_ + } =\int_x^1 {dy}f(\frac{x}{y}). \\ 
 \frac{4}{3}\bigg(\frac{{1 + {y^2}}}{{1 - y}}\bigg)
  - f(x)\int_0^1 {dy}. \frac{4}{3}\bigg(\frac{{1 + {y^2}}}{{1 - y}}\bigg) = {I_1} + {I_2}. \\ 
\end{gathered}
\label{2-5}‎
\end{eqnarray}
In the left integrand of (‎\ref{2-5}), an extra $y$ is placed to have the more useful parton momentum distribution function. Note that the DGLAP equation can be written for momentum distribution; i.e., in it ‎$ f(x) \to xf(x) $,‎ by putting an extra x on both sides, and a factor of ‎$ y.\frac{1}{y} = 1 $‎  within the integral. Different techniques of integration may be used in DGLAP. Here, we use only integration by parts:
\begin{eqnarray}
\begin{gathered}
u = f(\frac{x}{y}) \Rightarrow du = \frac{{df(\frac{x}{y})}}{{dy}}dy,  \\ dv = \frac{4}{3}(\frac{{1 + {y^2}}}{{1 - y}})dy \Rightarrow v = \int_{}^{} {dv}; \\ 
 {I_1} = f(x) v(1) - f(1) v(x) - \int_x^1 {v(y) \frac{{df(\frac{x}{y})}}{{dy}}dy}  \\ 
  = f(x) v(1) + \int_x^1 {v(\frac{x}{y})\frac{{df(y)}}{{dy}}dy},  \\ 
 {I_2} =  - f(x) [v(1) - v(0)] \\
 \Rightarrow I = f(x) v(0) + \int_x^1 {v(\frac{x}{y})\frac{{df(y)}}{{dy}}dy.}
\end{gathered}
\label{2-6}‎
\end{eqnarray}
Here and later, when convenient, the variable change ‎$ \frac{x}{y} \to y $‎ is used.  Bringing in a finite difference approximation of the differentials:\\
\begin{eqnarray}
\begin{gathered}
 I = v(0)f({x_i}) + \sum\limits_{k = 1}^i {\int\limits_{{x_k}}^{{x_{k - 1}}} {v({x_i}/y)\frac{{{f_k} - {f_{k - 1}}}}{{{x_k} - {x_{k - 1}}}}dy} }  \\ 
  = v(0)f({x_i}) + \int\limits_{{x_i}}^{{x_{i - 1}}} {v({x_i}/y)\frac{{{f_i} - {f_{i - 1}}}}{{{x_i} - {x_{i - 1}}}}dy}  \\ 
  + \sum\limits_{k = 1}^{i - 1} {\int\limits_{{x_k}}^{{x_{k - 1}}} {v({x_i}/y)\frac{{{f_k} - {f_{k - 1}}}}{{{x_k} - {x_{k - 1}}}}dy} }  \\
    = v(0)f({x_i}) + \frac{{{f_i}}}{{{x_i} - {x_{i - 1}}}}\int\limits_{{x_i}}^{{x_{i - 1}}} {v({x_i}/y)dy - \frac{{{f_{i - 1}}}}{{{x_i} - {x_{i - 1}}}}} \\ 
 { \int\limits_{{x_i}}^{{x_{i - 1}}} {v({x_i}/y)dy} } 
  + \sum\limits_{k = 1}^{i - 1} {\int\limits_{{x_k}}^{{x_{k - 1}}} {v({x_i}/y)\frac{{{f_k} - {f_{k - 1}}}}{{{x_k} - {x_{k - 1}}}}dy.} }  \\ 
\end{gathered}\label{2-7}‎
\end{eqnarray}
 Now, a bdd l.t. ‎${P_{qq}} $‎ can be read (extracted) from the coefficients of ‎$ f_k $‎ .
\begin{eqnarray}
{P_{qq}} : \left\{ \begin{gathered}
 {({P_{qq}})_{ii}} = v(0) + \frac{1}{{{x_i} - {x_{i - 1}}}}\int\limits_{{x_i}}^{{x_{i - 1}}} {v({x_i}/y)dy,}  \\ 
 {({P_{qq}})_{ik}} = \frac{1}{{{x_k} - {x_{k - 1}}}}\int\limits_{{x_k}}^{{x_{k - 1}}} {v({x_i}/y)dy}  \\ 
 -  \frac{1}{{{x_{k + 1}} - {x_k}}}\int\limits_{{x_{k + 1}}}^{{x_k}} {v({x_i}/y)dy.}  \\ 
\end{gathered} \right.
\label{2-8}‎
\end{eqnarray}

The other three kernels are essentially obtained in a similar manner and presented in Table \ref{I}, whose third column contains a small example set of numerical results whose $x-$points set is of actual use for NMC \cite{NMC}.

\begin{center}
\begin{table*}
\begin{tabular}{|l|c|c|}
\hline
Splitting Function & commuting matrices & Numerical‎ \\ \hline \hline
${P_{qq}} = \frac{4}{3}{(\frac{{1 + {y^2}}}{{1 - y}})_ + } $‎ & $\begin{array}{l}
 {({P_{qq}})_{ii}} = {v_{qq}}(0) + \frac{1}{{{x_i} - {x_{i - 1}}}}\int\limits_{{x_i}}^{{x_{i - 1}}} {{v_{qq}}({x_i}/y)dy}  \\ 
 {({P_{qq}})_{ik}} = \frac{1}{{{x_k} - {x_{k - 1}}}}\int\limits_{{x_k}}^{{x_{k - 1}}} {{v_{qq}}({x_i}/y)dy - \frac{1}{{{x_{k + 1}} - {x_k}}}\int\limits_{{x_{k + 1}}}^{{x_k}} {{v_{qq}}({x_i}/y)dy} }  \\ 
 \end{array}$ & $\begin{array}{l}
 \{  - {\rm{3}}.{\rm{96}},{\rm{2}}.{\rm{54}}, \\ 
 0.{\rm{62}},0.{\rm{29}}, \\ 
 0.{\rm{16}},0.{\rm{1}}0, \\ 
 0.0{\rm{6}},0.0{\rm{4}}\}  \\ 
 \end{array}$
 \\ \hline
‎${P_{qg}} = \frac{1}{2}[{y^2} + (1 - {y^2})] $‎ & $\begin{array}{l}
 {({P_{qg}})_{ii}} = {v_{qg}}(1) + \frac{1}{{{x_i} - {x_{i - 1}}}}\int\limits_{{x_i}}^{{x_{i - 1}}} {{v_{qg}}({x_i}/y)dy}  \\ 
 {({P_{qg}})_{ik}} = \frac{1}{{{x_k} - {x_{k - 1}}}}\int\limits_{{x_k}}^{{x_{k - 1}}} {{v_{qg}}({x_i}/y)dy - \frac{1}{{{x_{k + 1}} - {x_k}}}\int\limits_{{x_{k + 1}}}^{{x_k}} {{v_{qg}}({x_i}/y)dy} }  \\ 
 \end{array}$ & $\begin{array}{l}
 \{ 0.0{\rm{6}},0.0{\rm{7}}, \\ 
 0.0{\rm{4}},0.0{\rm{3}}, \\ 
 0.0{\rm{2}},0.0{\rm{2}}, \\ 
 0.0{\rm{1}},0.0{\rm{1}}\}  \\ 
 \end{array}$ \\ \hline
‎$ {P_{gq}} = \frac{4}{3}[\frac{{1 + {{(1 - y)}^2}}}{y}] $‎ & $\begin{array}{l}
 {({P_{gq}})_{ii}} = {v_{gq}}(1) + \frac{1}{{{x_i} - {x_{i - 1}}}}\int\limits_{{x_i}}^{{x_{i - 1}}} {{v_{gq}}({x_i}/y)dy}  \\ 
 {({P_{gq}})_{ik}} = \frac{1}{{{x_k} - {x_{k - 1}}}}\int\limits_{{x_k}}^{{x_{k - 1}}} {{v_{gq}}({x_i}/y)dy - \frac{1}{{{x_{k + 1}} - {x_k}}}\int\limits_{{x_{k + 1}}}^{{x_k}} {{v_{gq}}({x_i}/y)dy} }  \\ 
 \end{array}$ & $\begin{array}{l}
 \{ 0.{\rm{24}},0.{\rm{49}}, \\ 
 0.{\rm{56}},0.{\rm{63}}, \\ 
 0.{\rm{7}}0,0.{\rm{75}}, \\ 
 0.{\rm{79}},0.{\rm{82}}\}  \\ 
 \end{array}$ \\ \hline
‎$ \begin{array}{l}
 {P_{gg}} = \frac{{6y}}{{{{(1 - y)}_ + }}}  + 6y(1 - y) +\\  
 \\
 \frac{{6(1 - y)}}{y} + \delta (1 - y)(\frac{{33 - 2{n_f}}}{6}) \\ 
 \end{array} $‎ & $\begin{array}{l}
 {({P_{gg}})_{ii}} = {a_{gg}} + \frac{1}{{{x_i} - {x_{i - 1}}}}\int\limits_{{x_i}}^{{x_{i - 1}}} {{v_{gg}}({x_i}/y)dy}, \ {a_{gg}} =  - \frac{{33 + 2n_f}}{6} \\  
 {({P_{gg}})_{ik}} = \frac{1}{{{x_k} - {x_{k - 1}}}}\int\limits_{{x_k}}^{{x_{k - 1}}} {{v_{gg}}({x_i}/y)dy - \frac{1}{{{x_{k + 1}} - {x_k}}}\int\limits_{{x_{k + 1}}}^{{x_k}} {{v_{gg}}({x_i}/y)dy} }  \\ 
 
 \end{array}$ & $\begin{array}{l}
 \{  - {\rm{9}}.{\rm{43}},{\rm{6}}.{\rm{38}}, \\ 
 {\rm{2}}.{\rm{41}},{\rm{1}}.{\rm{9}}0, \\ 
 {\rm{1}}.{\rm{8}}0,{\rm{1}}.{\rm{8}}1, \\ 
 {\rm{1}}.{\rm{84}},{\rm{1}}.{\rm{88}}\}  \\ 
 \end{array}$ \\
 \hline
\end{tabular}
\caption{column 1: LO splitting functions. column 2: Commuting splitting function matrices, with $v_{ij}=\int_{}^{} {{P_{ij}}dy}$, for  $ij= qg$ or $gq$,  $v_{qq}\equiv {v}$ in (\ref{2-6}), and $v_{gg}= \frac{{6y}}{1 - y} + \frac{{6(1 - y)}}{y} + 6y(1 - y) $. Column 3: first column of the l.t., bdd splitting function matrices calculated from column 2, with a choice of $n = 8$, and ${x_1} = .712$.}
\label{I}
\end{table*}
\end{center}
\section{Commuting matrix Solution of DGLAP \\ 
 in \textit{X} Space for Finite $Q^2$ Interval}
The present section contains details of analysis of commuting matrix solutions of DGLAP equations ‎(\ref{1-1}) and (‎\ref{1-4}), for finite $Q^2$ interval. 

‎\subsection{Non-singlet}‎
 Using discrete form of the splitting function (\ref{2-8}),‎ discrete form of (‎\ref{1-1})‎ is:
 \begin{equation}\label{3-2}
\frac{{d{f_i^{NS}}(t)}}{{dt}} = \sum\limits_{j = 1}^i {{P_{ij}}f_j^{NS}(t) \Leftrightarrow \dot f(t) = P.f(t)} .
\end{equation}
 Independence of the kernel 
 ${P}={P_{qq}}$‎ from the variable ‎\textit{t} leads to the following solution to the matrix equation ‎(\ref{3-2})‎ for finite evolution from ${t_0}$ to $t$  
 \begin{equation}\label{3-3}
f(t) = {e^{(t - {t_0})P}}f({t_0}) \equiv {P_t}f({t_0}).
\end{equation}  
 To obtain the last equality in (\ref{3-3}), the diagonal and non-diagonal elements of the l.t. matrix ‎\textit{P}‎ can be separated into two commuting matrices as $P = {P_A} + {p_0}I$, with obvious definitions, including ${p_0} = {({P_{qq}})_{ii}}$.  Thus, being strictly l.t., ${P_A}^n = 0$ , and this in turn implies a finite expansion of the exponential as: 
\begin{equation}\label{3-4}	
 	{P_t}(t - {t_0}) = {e^{(t - {t_0}){P_0}}}I.\sum\limits_{k = 0}^{n - 1} {\frac{{{{({P_A})}^k}}}{{k!}}}. 
\end{equation}  	
Therefore, the finite evolution machinery, for the non-singlet solution of DGLAP, is constructed with the final $(t - {t_0})$ dependent, ready for computation, ${P_t}$ matrix.

‎\subsection{Singlet}

DGLAP evolution equation has the form of two coupled equations for the ‎\textit{singlet} quark distribution ‎$‎\Sigma‎$‎, coupled to gluons:\newline
\begin{eqnarray}
\begin{gathered}
‎\frac{\partial}{\partial t}‎\begin{pmatrix}
  ‎\Sigma‎(x,t)\\
  g(x,t)
\end{pmatrix} =‎ \hfill \\ \hfill \\
 \int_{x}^{1}‎\frac{dy}{y}
‎\begin{pmatrix}
P_{qq}(\frac{x}{y}) & 2n_fP_{qg}(\frac{x}{y})\\
P_{gq}(\frac{x}{y}) & P_{gg}(\frac{x}{y})
\end{pmatrix}\begin{pmatrix}
 ‎\Sigma‎(y,t)\\
  g(y,t)
\end{pmatrix}.
\end{gathered}
‎\label{1-4}‎
\end{eqnarray}
(\ref{1-4}) can be cast into a matrix equation as:
\begin{equation}\label{3-5}
	\frac{{\partial{f^S}(x,t)}}{\partial{t}} = \int_x^1 {\frac{{dy}}{y}{P_S}} (\frac{x}{y}){f^S}(y,t).
\end{equation}	

	Operating on the $2n$ dimensional external product space of the coupled pdfs ${f^S}$, discrete form of splitting functions are placed in a $(2n) \times (2n)$ matrix:
\begin{equation}\label{3-6}	
	{P_S} = \left( {\begin{array}{*{20}{c}}
   {\begin{array}{*{20}{c}}
   {{P_{qq}}} & {2{n_f}{P_{qg}}}  \\
\end{array}}  \\
   {\begin{array}{*{20}{c}}
   {{P_{gq}}} & {{P_{gg}}}  \\
\end{array}}  \\
\end{array}} \right).
\end{equation}	
 Based on arguments similar to those for (\ref{3-2}), (\ref{3-7}) gives a construction of the matrix equation of (\ref{3-5}), leading to its finite $(t-t_0)$ dependent solution, as in the case of (\ref{3-3}), due to independence of the kernel from ${t}$.  
\begin{equation}\label{3-7}	
{\dot f^S}(t) = {P_S}.{f^S}(t) \Rightarrow {f^S}(t) = {e^{(t - {t_0}){P_S}}}{f^S}({t_0}).
\end{equation}	
The kernel ${P_S}$ is expressed in an external (Kronecker or direct) product, $ \otimes $, space of a $2 \times 2$ matrix space and an $n \times n$ matrix space of bdd, l.t., thus commuting, splitting functions. ${P_S}$ is expanded, as a direct product, in terms of the four $2 \times 2$ identity and Pauli matrices: 
 \[{\sigma _1} = \left( {\begin{array}{*{20}{c}}
   {\begin{array}{*{20}{c}}
   0 & 1  \\
\end{array}}  \\
   {\begin{array}{*{20}{c}}
   1 & 0  \\
\end{array}}  \\
\end{array}} \right), \ {\sigma _2} = \left( {\begin{array}{*{20}{c}}
   {\begin{array}{*{20}{c}}
   0 & { - i}  \\
\end{array}}  \\
   {\begin{array}{*{20}{c}}
   i & 0  \\
\end{array}}  \\
\end{array}} \right), \ {\sigma _3} = \left( {\begin{array}{*{20}{c}}
   {\begin{array}{*{20}{c}}
   1 & 0  \\
\end{array}}  \\
   {\begin{array}{*{20}{c}}
   0 & { - 1}  \\
\end{array}}  \\
\end{array}} \right);\]
as follows:
\begin{eqnarray}
\begin{gathered}
{P_S} = {I}\otimes{P_0} + \vec P,  \\  \\
 \ {P_0} = \frac{1}{2}({P_{qq}} + {P_{gg}}), \\ \\
\vec P = {\sigma _3}\otimes{P_3} + \frac{i}{2}{\sigma _2}\otimes(2{n_f}{P_{qg}} - {P_{gq}}) + 
\frac{1}{2}{\sigma _1}\otimes
\\
(2{n_f}{P_{qg}} + {P_{gq}})
 = {\sigma _3}\otimes{P_3}+ 2{n_f}{\sigma _+}\otimes{P_{qg}} + {\sigma _-}\otimes{P_{gq}}, \\ \\ 
{P_3} = \frac{1}{2}({P_{qq}} - {P_{gg}}), 
 \end{gathered}
\label{3-9}‎
\end{eqnarray}
where ${\sigma _ \pm } = \frac{1}{2}({\sigma _1} \pm i{\sigma _2})$. Thus,
\begin{equation}\label{3-10}
	 {e^{(t - {t_0}){P_S}}} = {e^{(t - {t_0}){I}\otimes{P_0}}}{e^{(t - {t_0})\vec P}}. \\ 
 \end{equation}

 (\ref{3-10}) holds as the commutator $[{P_0}I,\vec P] = 0$, as an instance of easily proved $[I \otimes a, A \otimes b]=0$ for the direct product of $m \times m$ identity matrix $I$ and general matrix $A$ with any two bdd lower (or upper) triangular, thus commuting, $n \times n$  matrices $a$ and $b$.

Equation (\ref{3-10}) may be written in its matrix form, ready for computation:
\begin{eqnarray}
\begin{gathered}
 {e^{(t - {t_0}){P_S}}} =  {e^{(t - {t_0}){I}\otimes{P_0}}}(\sum\limits_{n = 0}^\infty  {\frac{{{{((t - {t_0})\vec P)}^n}}}{{n!}}} ) =  \\ 
 {{I}\otimes{e^{(t - {t_0}){P_0}}}}[{I}\otimes{Cosh((t - {t_0})\bar P)} +\\ 
 {I}\otimes{Sinh((t - {t_0})\bar P)} {{\bar P}^{ - 1}}\vec P].  
\end{gathered}
\label{3-11}‎
\end{eqnarray}

In (\ref{3-11}), $\bar P \equiv{\sqrt{{\vec P}^2}}$, is defining the length of $\vec P$, like a 2-norm, except that its components (vector coordinates) are not numbers but commuting bdd, l.t. matrices made of splitting functions. ${\bar P}^2 \equiv {\vec P}^2$ is a system of  
equations to be solved for computing $\bar P$.The last term in (\ref{3-11}) can be put in an alternative form: $Sinh((t - {t_0}){\vec{ P}})$.  Separating the diagonal and the strictly l.t. part of the splitting functions $P_0$ and $\bar P$ in (\ref{3-11}), brings the possibility of having a finite sum for the analytical evaluation of the singlet during the computation, similar to what was witnessed for the non-singlet, (\ref{3-4}). 

Equations (\ref{3-4}) and (\ref{3-11}) are called analytical commuting matrix solutions as they are analytically derived commuting matrices (finite sum representations of exponentials of commuting bdd l.t. matrices - computed numerically) for finite $Q^2$ interval solutions. In contrast, to bypass interpolation in data analysis in \cite{II}, a non-commuting matrix solution is formulated, using existing numerical routines for computation of exponentials of non-banded, thus non-commuting, triangular matrices.
 
\section{Comparison of pdf results for finite evolution of \textit{NS-S-}$g$ with MSTW and discussion}

\begin{figure}[ht]
\includegraphics[height=.23\textheight, width=1.\columnwidth]{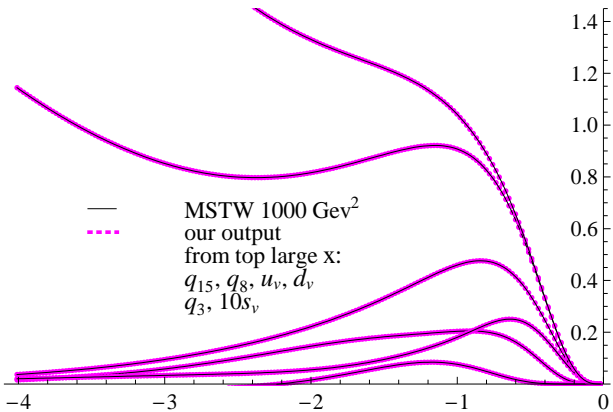}
\caption{Comparison of \textit{six} evolved valence and valence dominated  \textit{non-singlet} pdfs with MSTW, vs. $log_{10}(x)$, at $1000 Gev^2$. }
\label{1}
\end{figure}

\begin{figure}[ht]
\includegraphics[height=.23\textheight, width=1.\columnwidth]{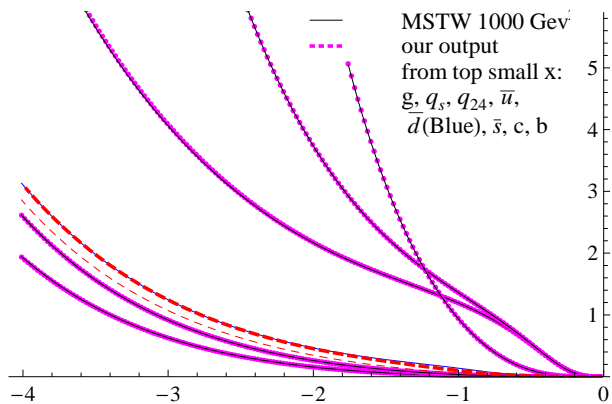}
\caption{Comparison of \textit{gluon} and \textit{four} sea or  sea influenced evolved pdfs:  \textbf{\textit{b, c}},  \textit{singlet}, and $q_{24}$ with MSTW, vs. $log_{10}(x)$, at $1000 Gev^2$.  \textit{Three} evolved pdfs: $ \bar{\textbf{\textit{u}}}, \bar{\textbf{\textit{d}}}, \bar{\textbf{\textit{s}}}$, are not compared due to triviality, see FIG. \ref{5} and explanation in the text.}
\label{2}
\end{figure}

The machinery thus far has equipped us with the capability to take a discrete set the initial or input pdfs, at x-points patterned according to (\ref{2-1}), at ${t_0}$ from some present standard LO set; evolve it ourselves to $t$ (our output), and then compare our results with their (final) pdfs at $t$. This is actually done via a quick computation, FIG. \ref{1}, \ref{2} with a small $n=180$, $x_1=.95$, (\ref{2-1}), input taken at $Q_0^2 = 10 {Gev^2}$  
from MSTW, \cite{MSTW}.

Our pdfs are assumed to be everywhere in the limit of high energy zero quark mass $SU_F(5)$ symmetry. Five flavor \textit{non-singlets} and \textit{singlet}, shown in FIG. \ref{1}, \ref{2}, have definitions:
\begin{eqnarray}
\begin{gathered}
 {q_{3}}=u-d, \hspace{3mm}  {q_{8}}=u+d-2s, \hfill \\
 {q_{15}}=u+d+s-3c, \hspace{3mm} {q_{24}}=u+d+s+c-4b, \hfill \\ 
{q_s}\equiv{\Sigma}=u+d+s+c+b; \hfill \\
 q\equiv{q_{tot}}+\bar{q}, \hspace{3mm} q=u,\cdots, b. \hfill 
\end{gathered}\label{4-1}‎\end{eqnarray}

In addition, three \textit{non-singlet valence} quarks are assumed along with MSTW: 
\begin{equation}
\label{4-2}
	  q_v\equiv{q_{tot}}-\bar{q}, \hspace{3mm} q=u, d, s; \hfill 
 \end{equation}
and are directly evolved; furthermore, all the sea quarks evolution is inferred from (\ref{4-1}, \ref{4-2} ) for the figures.
 
There are two further assumptions. It is essential to have a continuous coupling constant, $\alpha_s$, as flavor number changes, as $Q^2$ crosses the quark mass $m_b^2=22.56 \ Gev^2$, \cite{MSTW}, in the finite interval of evolution, $\delta{Q^2}=[10,1000]  \  Gev^2$. The LO value of $\alpha_s(m_z)=.13939$ of MSTW is used here together with the related considerations \cite{MSTW, MSTWalpha}. Absence of this continuity constraint, brings inconsistency and considerably increased deviations. We also assume a simple continuity of pdfs as the flavor number changes at the same $Q^2=m_b^2$. We are not claiming that there is no discrepancy between our assumptions and those of MSTW.

\begin{figure}[ht]
\includegraphics[height=.23\textheight, width=1.\columnwidth]{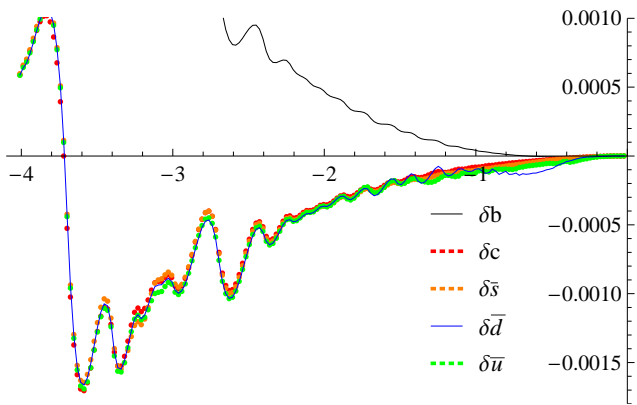}
\caption{Absolute deviations of the \textit{five} indirectly evolved \textit{sea} pdfs of FIG. \ref{2} from MSTW. 
}
\label{5}
\end{figure}

\begin{figure}[ht]
\includegraphics[height=.23\textheight, width=1.\columnwidth]{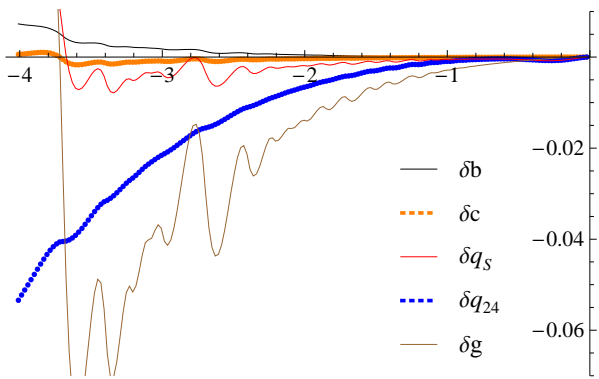}
\caption{Absolute deviations of \textit{five} evolved pdfs with respect to MSTW. It indicates a relation between the problems of large positive $\delta{\textbf{\textit{b}}}$ and large negative $\delta{q_{24}}$ posed in the text.}
\label{6}
\end{figure}

The quality of match of our pdf outputs at ${Q^2} = 1000 {Gev^2}$ (magenta dots in comparison) with MSTW (solid black lines) can be observed in FIG. \ref{1}, \ref{2}.
Similar matching with any other good solution of DGLAP equations at LO is expected. We have had explicit trials with CTEQ5L \cite{CTEQ} and GRV \cite{GRV} pdfs.
 In FIG. \ref{2}, \textit{three} indirectly evolved pdfs: $ \bar{\textbf{\textit{u}}}, \bar{\textbf{\textit{d}}}, \bar{\textbf{\textit{s}}}$, are not compared to MSTW. Their absolute deviations from MSTW pdfs are very small and very similar to that of the \textbf{\textit{c}} quark, while very different from the \textbf{\textit{b}} quark, FIG. \ref{5}, \ref{6}. Thus, their comparison is just as good as that of the \textbf{\textit{c}} quark. Close scrutiny of FIG. \ref{2} shows the difference of fit between the two, \textbf{\textit{c}} and \textbf{\textit{b}} quarks. 
 There will be a related discussion of \textbf{\textit{b}} problem in the ending paragraphs.

Quantitative description of the fit, depending on choice of the beginning point $x_1$, may be described, at each point $x$, (\ref{2-1}), in terms of absolute deviation, $\delta$, defined as deviation of our result from the corresponding (final) MSTW pdf; or relative deviation, $\delta_R$, defined as $\delta$ divided by the value of the (final) MSTW pdf. It can be seen that our method based on discrete $x$, (\ref{2-1}), induces oscillations in the deviations. 

\begin{figure}[ht]
\includegraphics[height=.23\textheight, width=1.\columnwidth]{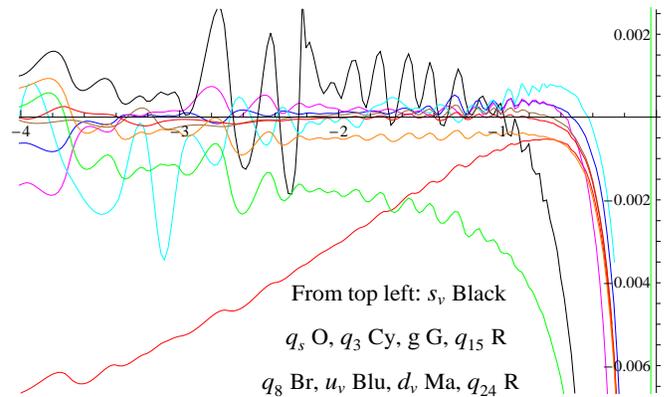}
\caption{Relative deviations of all directly evolved \textit{ nine NS-S-g} pdfs with respect to MSTW.}
\label{4}
\end{figure}

\begin{figure}[ht]
\includegraphics[height=.23\textheight, width=1.\columnwidth]{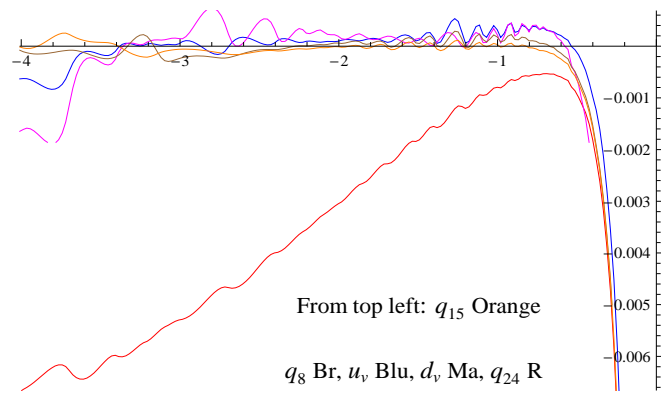}
\caption{Relative deviations of \textit{four} directly evolved pdfs with smallest, and \textit{one}, $q_{24}$, with largest maximum $\delta_R$ for $x\in(10^{-4}, \sqrt{10})$ (see text for $\sqrt{10})$).}
\label{3}
\end{figure}

\begin{figure}[ht]
\includegraphics[height=.23\textheight, width=1.\columnwidth]{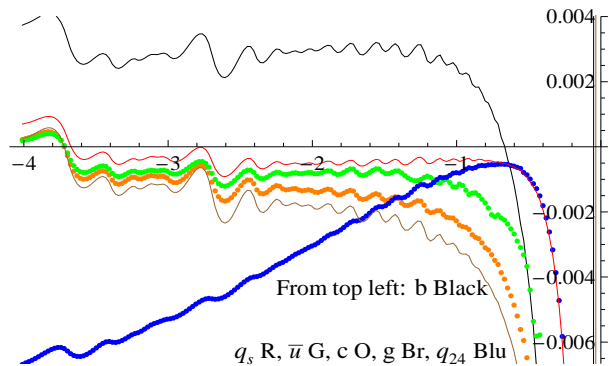}
\caption{Relative deviations of {six} evolved pdfs, related to FIG. \ref{6}. The place of $\delta_R{\bar{\textbf{\textit{d}}}}$ and $\delta_R{\bar{\textbf{\textit{s}}}}$ is between $\delta_R{\bar{\textbf{\textit{u}}}}$ and $\delta_R{\textbf{\textit{c}}}$. Large $\delta_R{\textbf{\textit{b}}}$ and $\delta_R{q_{24}}$ result from their counterparts in FIG. \ref{6}.}
\label{7}
\end{figure}

 The order of calculation of evolved pdfs is according to increasing $i$, or decreasing $x_i$, in (\ref{2-1}). As $x_i$ decreases, three areas may be differentiated. Relative deviations for 
 large $x$ area are very large, which is not far from expectation due to very small size of denominator pdfs there, and having largest finite difference intervals in $x$, without smoothening effects of summation, of the convolution integral, which is just beginning there.
 
  The second middle to small $x$ area approximately corresponds to $x\in(10^{-4}, \sqrt{10})$ (stricly speaking, for near zero gluon and $s_v$ quark there $ \sqrt{10}\longrightarrow 2\sqrt{10}$, FIG. \ref{4}). In this area, for $u_v, d_v, q_8, q_{15}$ which have the smallest relative deviations, FIG. \ref{4}, maximum of $\delta_R$, FIG. \ref{3}, is below $.0007$ (neglecting the last small $x$ oscillation of $d_v$).
   For $s_v$, FIG. \ref{4}, $\delta_R$ is somewhat larger and problematic, due to smallness of $s_v$, and that both input and output of MSTW cross zero in the middle of the second area. For $q_{24}$, $\delta_R$ is the largest, FIG. \ref{4}, by a considerable margin; its maximum is below$.007$, FIG. \ref{3}, at $x\approx 10^{-4}$. 
  It may be expected from application  of finite difference method, that deviations and maximum $\delta_R$ decrease with increase in $x_1$. We found this to be the case in a limited trial of letting $x_1=.95$ to take powers of $1/4, 1, 2$. 

  The exceptional behavior of $ q_{24}$ is related to different behavior of \textit{b} quark from the other \textit{sea} quarks, FIG. \ref{5}, \ref{6}, \ref{7}.  Further investigation is demanded; it may be due to difference of our assumptions and MSTW's, e.g., it may be related to flavor number scheme. 
  
  In the third area of deeper small $x$, approximately $x<10^{-4}$, $\delta_R$ ceases to be as good. We may return to details of the subject in future. 

In conclusion, we have developed a detailed numerical presentation of a commuting matrix solution of DGLAP evolution equations at LO PQCD, as a first step of realizing the potential of doing non-parametric data analysis. 
 
\section{acknowledgments}
MG gratefully acknowledges use of analytical and user friendly numerical tools of Mohammad Zandi, developed (2009-10) for \cite{II}, and pedagogical presence of Ali Mollabashi (2007) in this paper. 
This work was supported in part by the research funds of Shahid Beheshti University, G. C. 600/884 (1389-5-24).

 \thefootnote{*}Subject of this paper was well developed in the 3rd reference of \cite{Spin2000}, including a comparison with GRV \cite{GRV}, and discovery of singularity, an obstacle in the way of its further development, \cite{Spin2002, II}.

 \thefootnote{**}
In this paper, there are two uses for the symbol $\otimes$: convolution integral on the one hand; and "direct", "external" or "Kronecker" product of matrices on the other hand, depending on the context.  

\end{document}